\documentclass[aps,prl,twocolumn,showpacs,preprintnumbers,superscriptaddress]{revtex4}

\usepackage{amsmath}
\usepackage{amssymb}
\usepackage{graphicx}
\usepackage{dcolumn}
\usepackage{bm}
\newcommand{\YP}{\rm Yb_{3}Pt_{4}}
\newcommand{\YRS}{\rm YbRh_{2}Si_{2}}

\begin{document}

\title{Localized Moments and the Stability of Antiferromagnetic Order in Yb$_{3}$Pt$_{4}$}

\author{L. S. Wu}
\affiliation{Department of Physics and Astronomy,
Stony Brook University, Stony Brook, New York 11794-3800, USA}
\author{Y. Janssen}
\affiliation{Condensed Matter Physics and Materials Science
Department, Brookhaven National Laboratory, Upton, New York
11973-5000, USA}
\author{M. C.  Bennett}
\affiliation{Condensed Matter Physics and Materials Science
Department, Brookhaven National Laboratory, Upton, New York
11973-5000, USA}
\author{M. C. Aronson}
\affiliation{Department of Physics and Astronomy,
Stony Brook University, Stony Brook, New York 11794-3800, USA}\affiliation{Condensed Matter Physics and Materials Science
Department, Brookhaven National Laboratory, Upton, New York
11973-5000, USA}
\date{\today}

\begin{abstract}

We present here the results of electrical resistivity $\rho$, magnetization \emph{M}, ac susceptibility $\chi^{\prime}_{ac}$, and specific heat $C_{\rm M}$ measurements that have been carried out on single crystals of $\YP$ over a wide range of fields and temperatures. The 2.4 K N\'{e}el temperature that is found in zero field collapses under field to a first order transition $T_{\rm N}=0$ at $B_{\rm CEP}=1.85$ T. In the absence of antiferromagnetic order, the specific heat $C_{\rm M}(T,B)$, the magnetization $M(T,B)$, and even the resistivity $\rho(T,B)$ all display $B/T$ scaling, indicating that they are dominated by strong paramagnetic fluctuations, where the only characteristic energy scale results from the Zeeman splitting of an energetically isolated, Yb doublet ground state. This paramagnetic scattering disappears with the onset of antiferromagnetic order, revealing Fermi liquid behavior $\Delta\rho=AT^{2}$ that persists up to the antiferromagnetic phase line $T_{\rm N}(B)$, but not beyond. The first order character of $T_{\rm N}=0$ and the ubiquity of the paramagnetic fluctuations imply that non-Fermi liquid behaviors are absent in $\YP$. In contrast to heavy fermions like $\YRS$, $\YP$ represents an extremely simple regime of $f$-electron behavior where the Yb moments and conduction electrons are almost decoupled, and where Kondo physics plays little role.
\end{abstract}

\pacs{75.30.Kz, 75.50.Ee, 75.40.Cx, 71.20.Eh}

\maketitle

\section{Introduction}

The quantum critical point (QCP) that is formed when magnetic order is suppressed to zero temperature is firmly established as an integral part of the physics of most strongly correlated electronic materials. Arguably, the most comprehensive account of these phenomena comes from studies of heavy fermion compounds~\cite{coleman2005,stewart2001,vonlohneysen2007,vonlohneysen2008,gegenwart2008,steglich2010}. Initially, it was thought that the unusual divergencies of the specific heat and magnetic susceptibilities that were found near QCPs, as well as electrical resistivities with linear temperature dependencies, phenomena collectively referred to as `non-Fermi liquid' behavior, reflected the dominance of quantum critical fluctuations. However, it has become clear that in at least a few cases, that the QCP affects the electronic structure itself, where $T=0$ electronic delocalization leads to a change in the Fermi surface volume at or near the QCP~\cite{coleman2001, si2001, senthil2003}. Evidence for these Fermi surface volume changes come from Hall effect measurements near the field-driven QCP in $\YRS$~\cite{paschen2004,friedemann2009,friedemann2010}, from discontinuous changes in quantum oscillations and moment localization near the pressure - driven QCP in CeRhIn$_{5}$~\cite{shishido2005,park2008}, and from the values of the quantum critical exponents themselves~\cite{si2001,schroder2000}.

$\YRS$ exemplifies the full range of phenomena that can be associated with a field-driven QCP~\cite{gegenwart2008}.  First, the $B=0$ N\'{e}el temperature is only 0.065 K, with a correspondingly small ordered moment $\simeq$ 10$^{-3}\mu_{B}$/Yb. $T_{\rm N}$ is suppressed continuously to $T_{\rm N}=0$ with a field \mbox{$B=0.66$ T}~\cite{trovarelli2000,gegenwart2002}. Quantum critical scaling of the field and temperature dependencies of the specific heat $C$ and the magnetization $M$ are reported in the vicinity of the QCP~\cite{custers2003}. non-Fermi liquid temperature dependencies are observed near the QCP, such as a diverging specific heat $C/T\simeq$-ln($T$), and a linear temperature dependence for the electrical resistivity \mbox{$\rho(T)=\rho_{0}+aT$}, observed over several decades in temperature~\cite{trovarelli2000,custers2003}. Fermi liquid behavior is found once the antiferromagnetic order is suppressed by fields $B\geq B_{\rm QCP}$, with $\rho=\rho_{0}+A(B)T^{2}$ and $C=\gamma(B)T$. The Fermi liquid parameters $A$ and $\gamma$ indicate that the quasiparticle mass is strongly enhanced and even diverges as $B\rightarrow B_{\rm QCP}$ from above, signalling the breakdown of the Fermi liquid itself at the QCP. Associated with this breakdown is an electronic localization transition, where the number of states contained by the Fermi surface changes  at or near the QCP~\cite{paschen2004,gegenwart2007,gegenwart2008,friedemann2009,friedemann2010}.

The question that we ask here is what part of  this spectrum of quantum critical phenomena survives in a more minimal system, where electronic localization does not occur. $\YP$ is an ideal system in which to explore this issue. Metallic $\YP$  orders antiferromagnetically at $T_{\rm N}=2.4$ K~\cite{bennett2009}, where the mean-field like development of the ordered parameter taken from neutron diffraction measurements results in a T=0 moment of 0.8 $\mu_{B}$/Yb~\cite{janssen2010}. Specific heat and inelastic neutron scattering measurements indicate that the antiferromagnetic order develops from Yb moments in a crystal field split doublet ground state that is well separated in energy from the first excited state~\cite{aronson2010}. The rapid recovery of the magnetic entropy $S(T_{\rm N})=0.8$ Rln2 suggests that there is little evidence that Kondo compensation of the Yb moments has occurred as $T\rightarrow T_{\rm N}$, indicating that $T_{\rm K}\leq T_{\rm N}$. For these reasons, it is believed that the Yb moments in $\YP$ are spatially localized, and only weakly coupled to the conduction electrons. Given the apparent irrelevance of Kondo physics to $\YP$, it likely that the 4$f$-holes of the Yb ions are excluded from the $B=0$ Fermi surface. The complexity of the unit cell in $\YP$ precludes a direct test of this conclusion from electronic structure calculations.

Magnetic fields suppress antiferromagnetic order in $\YP$,  and we find that $T_{\rm N}=0$ for the critical end point (CEP) $B_{\rm CEP}=1.85$ T. The Clausius - Clapeyron equation is obeyed here, and although the antiferromagnetic phase line intersects the $T=0$ axis vertically and so cannot be fitted to a power-law as $T_{\rm N}\rightarrow0$, $T_{\rm N}(B)$ is continuous for $T_{\rm N}>0$~\cite{wu2011}, possibly following a mean-field expression. We present here the results of experiments that seek answers to three questions. First, is there non-Fermi liquid behavior near $B_{\rm CEP}$ in $\YP$?  Measurements of the temperature dependence of the electrical resistivity are expected to be of particular importance in answering this question.  Second, does a Fermi liquid state develop once magnetic fields suppress antiferromagnetic order?  If so, is there a divergence of the Sommerfeld coefficient $\gamma$ and the resistivity coefficient $A$ as $B\rightarrow B_{\rm CEP}$ that signal the breakdown of this Fermi liquid with the onset of antiferromagnetic order?  Finally, is there any suggestion of electronic delocalization in $\YP$, or is the coupling between the Yb moments and the conduction electrons always vanishingly small?  Electrical resistivity, specific heat, magnetic susceptibility, and magnetization measurements were performed on $\YP$ over a wide range of fields and temperatures, in both the antiferromagnetic and paramagnetic phases. The results are presented here, and are compared to those of similar experiments on $\YRS$, with the intention of providing support for a global phase diagram that relates these two very different systems.

\section{Experimental Details}
Single crystals of $\YP$ were grown from lead flux\cite{bennett2009,janssen2010}, yielding rod-like crystals with approximate dimensions of  $1\times1\times5$ mm. Powder X-ray diffraction measurements were used to verify that the crystals form in the reported rhombohedral Pu$_{3}$Pd$_{4}$ structure type\cite{palenzona1977}. Electrical resistivity $\rho$ was measured using a Quantum Design Physical Property Measurements System (PPMS) for temperatures T between 0.1 K and 300 K, and in fields as large as 6 T. Electrical contacts were made to the crystal in the four probe configuration using silver-filled epoxy. The current flowed along the long axis of the crystal, corresponding to the crystallographic \emph{c-axis}.  The dc magnetization M was measured using a Quantum Design Magnetic Property Measurement System (MPMS) in magnetic fields $B$ as large as 7 T. Measurements of the ac magnetic susceptibility $\chi_{ac}^{\prime}$ were also carried out in the MPMS using a 17 Hz ac field $B_{\rm ac}=4.17$ Oe, with an additional dc field that was as large as 2 T. Specific heat $C$ was measured using the PPMS for temperatures $T$ between 0.3 K and 300 K, and in fixed magnetic fields $B$ that were as large as 7 T. All the measurements that we report here were were carried out with the magnetic field perpendicular to the \emph{c}-axis, and due to the small anisotropy within the easy \emph{ab} plane, the field direction within the \emph{ab} plane is not specified.

\section{Experimental Results }

\begin{figure}
\includegraphics[width=6.0cm]{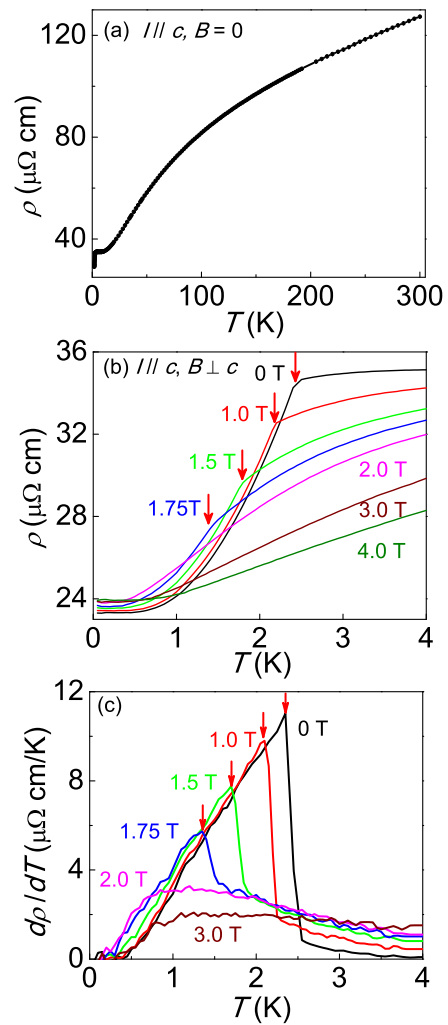}
\caption{(Color online) (a) The temperature dependence of the electrical resistivity $\rho(T)$ in $\YP$. (b) The temperature dependencies of $\rho(T)$ measured in different magnetic fields from 0 T to 4.0 T, as indicated. The red arrows indicate the antiferromagnetic transitions at each field \mbox{$B\leq 1.85$ T}. (c) The temperature derivative of the electrical resistivity $d\rho/dT$ in different fixed fields, as indicated. Red arrows indicate values of $T_{\rm N}(B)$, taken from the maxima in $d\rho/dT$. \label{RvsT}}
\end{figure}

\begin{figure}
\includegraphics[width=7.50cm]{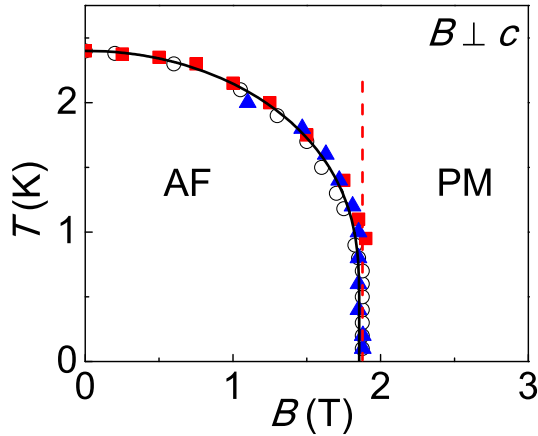}
\caption{(Color online)  Field - temperature phase diagram of $\YP$. The antiferromagnetic ordering temperatures $T_{\rm N}(B)$ extracted from the temperature(red filled squares) and field (blue triangles) dependent resistivities are in good agreement with the phase line determined from specific heat(black open circles) measurements in \cite{wu2011}. Solid black line is a fit to a mean-field expression. The vertical dashed line indicates the critical field $B_{\rm CEP}=1.85$ T, above which no antiferromagnetic order is found.  \label{BT}}
\end{figure}

Electrical resistivity has proven to be a very sensitive probe of the quantum critical fluctuations in other heavy fermion compounds where antiferromagnetic order can be suppressed to $T=0$~\cite{rosch1999,stewart2001}. The temperature dependence of the $B=0$ electrical resistivity $\rho(T)$ in $\YP$ is shown in Fig.~\ref{RvsT}a. $\rho(T)$ drops monotonically from its room temperature value of 127 $\mu\Omega$-cm to 35 $\mu\Omega$-cm at 10 K, confirming that $\YP$ is definitively metallic. Given the crystal field scheme deduced from specific heat and inelastic neutron scattering measurements where four doublets are separated by 87 K, 244 K, and 349 K~\cite{aronson2010}, it is likely  that the bulge in $\rho(T)$ at intermediate temperatures reflects the depopulation of these crystal field levels with reducing temperature. The onset of antiferromagnetic order is evident from the sharp drop in $\rho(T)$ at the N\'{e}el temperature $T_{\rm N}=2.4$ K. Since our primary interest is in the behavior of $\rho(T)$ as magnetic fields suppress $T_{\rm N}$ to zero, we have repeated the measurements of $\rho(T)$ in different fixed fields B ranging from 0 T to 4 T (Fig.~\ref{RvsT}b). As expected, the resistive drop at $T_{\rm N}$ occurs at lower temperatures with increasing fields, and there is no indication of a resistive anomaly when $B\gtrsim2$ T. We take $T_{\rm N}(B)$ from the maximum in the temperature derivative, $d\rho/dT$ (Fig.~\ref{RvsT}c), and the result is compared in Fig.~\ref{BT} to the phase line $T_{\rm N}(B)$ that was previously determined from specific heat, neutron diffraction, and magnetization measurements~\cite{wu2011}. We note that the specific heat measurements place $B_{\rm CEP}$ near 1.9 T, although the other measurements find $B_{\rm CEP}\simeq1.8-1.85$ T. The agreement is very good, especially considering that the experiments were performed on different crystals, and that small uncertainties in the orientation of the field are inevitable. We will take $B_{\rm CEP}=1.85\pm0.05$ T.

\begin{figure}
\includegraphics[width=7.0cm]{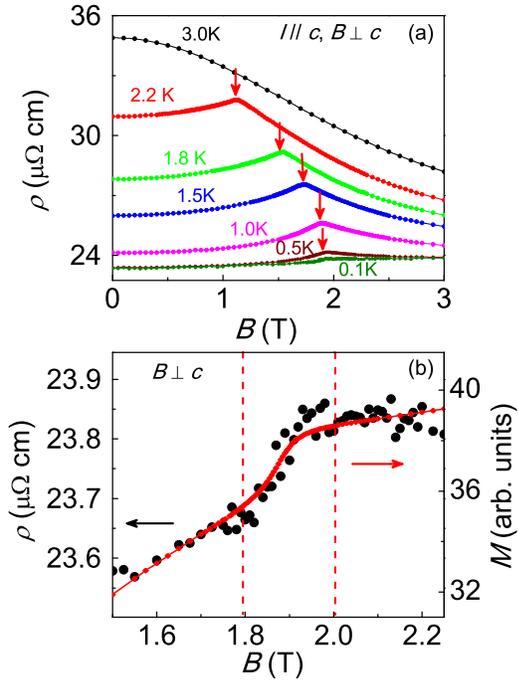}
\caption{(Color online) (a) Field dependencies of the electrical resistivity $\rho$ measured at different temperatures from 0.1 K to 3.0 K, as indicated. Red arrows indicate the antiferromagnetic transitions. (b) Magnetoresistivity $\rho$ ($\bullet$, left axis) measured at 0.1 K plotted together with the magnetization (red solid line, right axis) measured at 0.2 K.  Vertical dashed lines delineate the step like kink around the critical field $\thicksim 1.85 $ T. \label{RvsB}}
\end{figure}

Since the phase line is very steep when $T_{\rm N}\rightarrow0$, measurements of the field dependence of the resistivity at different fixed temperatures are better suited to exploring this part of the $T-B$ phase diagram. As indicated in Fig.~\ref{RvsB}a, the magnetoresistance $\rho$(B) has a sharp peak at $T_{\rm N}(B)$, most prominent for $T_{\rm N}\geq1$ K. The values of $T_{\rm N}(B)$ that are taken from this peak have been added to the phase diagram in Fig.~\ref{BT}, and they agree well with those found from the $\rho(T)$ data of Fig.~\ref{RvsT}. At lower temperatures, the peak in $\rho(B)$ evolves into a broadened step, whose magnitude becomes smaller with decreasing temperature.  No hysteresis was observed between resistivity measurements obtained with increasing or decreasing fields, even at the lowest temperatures.  The values of $T_{\rm N}(B)$ that are taken from the resistive step have been added to Fig.~\ref{BT}, and we see that the magnetoresistivity data closely track the near-vertical phase line $T_{\rm N}(B)$ as it approaches the horizontal axis at the critical field $B_{\rm CEP}=1.85$ T. The width of the  field-induced step in $\rho(B)$ decreases with decreasing temperature, and at the lowest temperatures it has a width of $\simeq0.2$ T. This behavior is reminiscent of the step in the $\YP$  moment observed in both magnetization $M(B)$ and neutron diffraction measurements~\cite{wu2011}. We previously showed that the step $\Delta M$ and the vertical phase line $T_{\rm N}(B)$ are in agreement with the Clausius-Clapeyron equation, indicating that antiferromagnetic order in $\YP$  vanishes at a first order transition or critical endpoint,  where $T_{\rm N}=0$ and $B=B_{\rm CEP}$. We have compared the magnetization step measured at $T=0.2$ K to the magnetoresistivity step measured at 0.1 K in Fig.~\ref{RvsB}b, and their resemblance is striking. This is our first indication that the magnetization controls the electrical resistivity in $\YP$, a finding that we will develop further below.

\begin{figure}
\includegraphics[width=6cm]{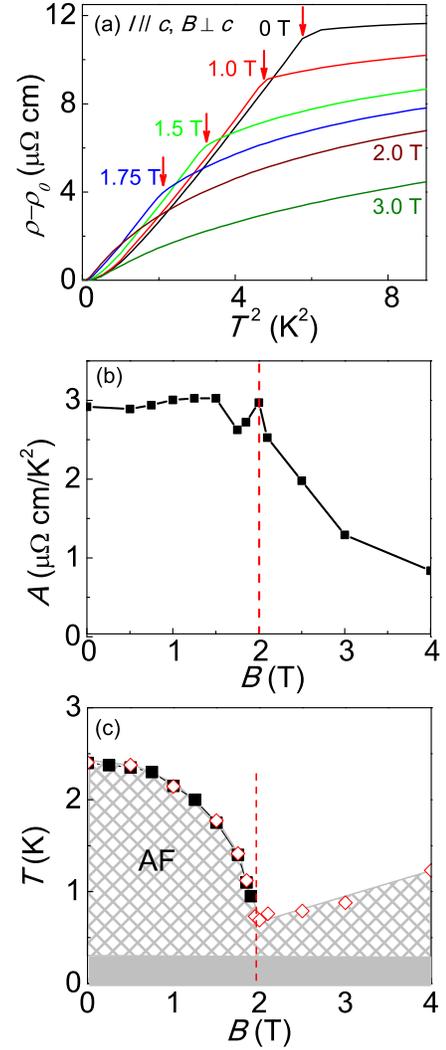}
\caption{(Color online) (a) The resistivity $\rho-\rho_{0}$ as a function of $T^2$ in different magnetic fields as indicated. Red arrows indicate the antiferromagnetic transitions, taken from the maxima in $d\rho/dT$. (b) The coefficient of the quadratic temperature dependence $A$ as a function of  magnetic field $B$. Vertical dashed line indicates the critical field $B_{\rm CEP}=1.85$ T. (c) The range of fields and temperatures $T\leq T_{\rm FL}(B)$ where $\rho-\rho_{0}=AT^{2}$ is indicated by the shaded areas on this field-temperature phase diagram. $T_{\rm FL}$ is indicated in the paramagnetic phase ($B\geq B_{\rm CEP}$) by red open diamonds. $T_{\rm N}(B)$ is indicated by the black squares, and the critical field $B_{\rm CEP}$  by the vertical dashed line. The lower temperature limit for our measurements is $\simeq$ 0.3 K. \label{dRdT}}
\end{figure}

The suppression of magnetic order in a heavy fermion compound that has been driven to a QCP often results in a normal metallic state that is a Fermi liquid. Here, the electrical resistivity is quadratic in temperature $\rho(T)=\rho_{0}+AT^{2}$, and the coefficient $A$ is often enhanced near the QCP, reflecting the growth of quasiparticle interactions that can culminate in the divergence of the quasiparticle mass at the QCP itself.  Accordingly, we have plotted the temperature dependent part of the electrical resistivity $\rho(\rm T)-\rho_{\rm 0}$, measured in different fixed fields,  as a function of $T^{2}$ in Fig.~\ref{dRdT}a.  A quadratic temperature dependence is observed within the antiferromagnetically ordered state, i.e. for $T\leq T_{\rm N}(B)$. There is only a small variation in the slopes of the curves in Fig.~\ref{dRdT}a for the fields $B\leq1.85$ T where antiferromagnetic order is present.  To highlight this point, we have plotted the coefficient $A(B)$ in Fig.~\ref{dRdT}b, and within the antiferromagnetic phase $A(B)$ remains roughly constant. We have attempted to extend the Fermi liquid temperature dependence to higher fields $B\geq B_{\rm CEP}$, but we find that the fit is only valid over an extremely small range of temperatures $T\leq T_{\rm FL}(B)$ and the minimum measurement temperature, 0.3 K (Fig.~\ref{dRdT}c). While we report $A(B)$ that is derived from these fits in Fig.~\ref{dRdT}b, we feel that there is no convincing evidence that Fermi liquid behavior can be detected in $\rho(T)$, once antiferromagnetic order has  been suppressed to zero by either temperature or field.

\begin{figure}
\includegraphics[width=6.0cm]{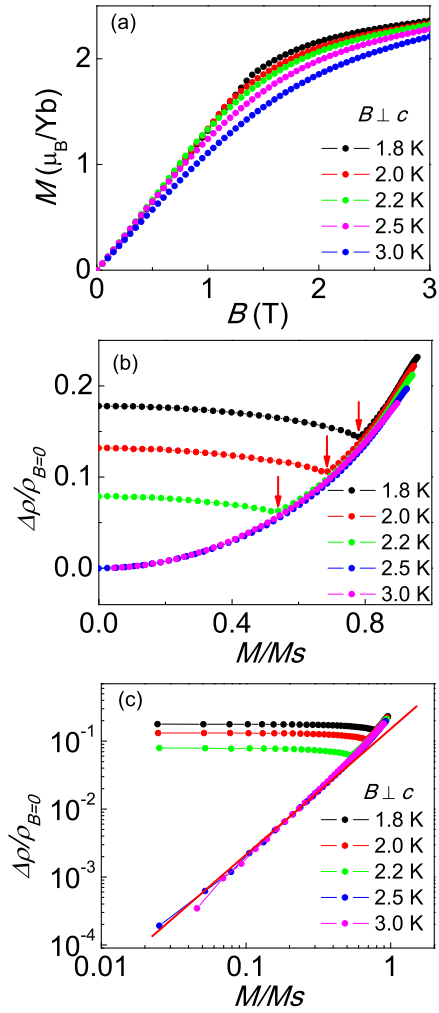}
\caption{(Color online) (a) The field dependencies of the  magnetization $M$, determined at different temperatures between 1.8 K to 3.0 K. (b) Plot of the normalized magnetoresistivities as functions of the normalized magnetizations $M/M_{\rm S}$ measured at different temperatures, where $\Delta\rho(B)=\rho(B)-\rho(B=0)$, and $M_{\rm S}$ is the saturation magnetization defined in the text. The red arrows indicate the onset of antiferromagnetic order for each curve, highlighting that this relationship fails within the antiferromagnetic phase.  (c) The data from (b) collapse onto a single curve with a slope of 2, as indicated by the red line. \label{SD}}
\end{figure}

If the paramagnetic state with $B\geq B_{\rm CEP}$ is not a Fermi liquid at low temperatures, then what physical processes are responsible for the electrical resistivity once antiferromagnetic order is suppressed? The similar field dependencies of the magnetoresistivity $\rho(B)$ and the magnetization $M(B)$ displayed in Fig.~\ref{RvsB}b suggest that spin disorder scattering may dominate. To test this idea, we have combined measurements of $M(B)$(Fig.~\ref{SD}a), normalized by $M_{\rm S}$, which is taken to be the value of $M$ for $T=1.8$ K and $B=3$ T, with those of the normalized magnetoresistivity $\Delta\rho/\rho(B=0) = (\rho(B)-\rho(B=0))/\rho(B=0)$. The result is presented in Fig.~\ref{SD}b. The normalized magnetoresistivity obtained at different fixed temperatures collapses as a function of the normalized magnetization, provided that the fields and temperatures of the respective measurements do not place $\YP$ within the antiferromagnetic phase, whose boundaries are indicated by arrows in Fig.~\ref{SD}b. Spin-disorder scattering can be identified by its power-law relation between $\rho(T,B)$ and $M(T,B)$,  where $\Delta\rho/\rho(B=0)\propto (1-(M/M_{\rm S})^{2})$~\cite{degennes1958,fisher1968}. This relationship is confirmed in Fig.~\ref{SD}c, where a double logarithmic plot of $\Delta\rho/\rho(B=0)$ is linear with respect to $M/M_{\rm S}$. The best fit to the scaling region  gives a slope of two, as indicated by the red line. Our measurements affirm our proposal that fluctuations in the magnetization are the primary agent for scattering quasiparticles over a very wide swath of the $B-T$ phase diagram, provided that antiferromagnetic order is not present.

\begin{figure}
\includegraphics[width=6.0cm]{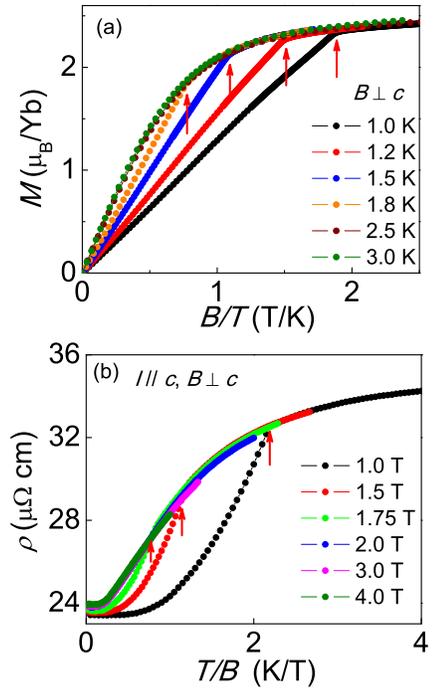}
\caption{(Color online) (a) Field dependencies of the magnetization $M$ were measured at different temperatures, and then plotted as functions of $B/T$. (b) The temperature dependencies of the electrical  resistivity $\rho$ were measured in different magnetic fields from 1.0 T to 4.0 T, and then plotted as functions of $T/B$. Red arrows indicate the onset of antiferromagnetic order, showing that the resistivity data collapse onto a single curve in the paramagnetic phase. \label{ScalingM}}
\end{figure}

A simple scaling analysis reveals the nature of the dominant magnetization fluctuations. Fig.~\ref{ScalingM}a shows that the magnetization $M$ collapses when plotted as a function of $B/T$, but only when $B$ and $T$ are taken from the paramagnetic part of the $(T,B)$ phase diagram (Fig.~\ref{BT}). Since Fig.~\ref{SD} shows that the magnetoresistivity is a proxy for the magnetization, it is not surprising that it too displays $B/T$ scaling(Fig.~\ref{ScalingM}b). This scaling fails within the antiferromagnetic phase $T\leq T_{\rm N}(B)$, where Fermi liquid behavior $\Delta\rho=AT^{2}$ is observed. The success of the $B/T$ scaling implies that the magnetization fluctuations are simply paramagnetic fluctuations among the crystal field split states of the Yb$^{3+}$ ion. The crystal field split manifold of the $J=7/2$ Yb$^{3+}$ ions in rhombohedral symmetry consists of four doublets, and inelastic neutron scattering and specific heat measurements indicate that the ground doublet in $\YP$ is separated from the first excited level by $~80-90$ K~\cite{janssen2010,aronson2010}, much larger than the temperature scales probed in the measurements reported here. Practically speaking, we can safely ignore the excited states, and so the field and temperature dependencies of the magnetization $M$ reflect the two-fold degeneracy of the ground doublet, lifted by Zeeman splitting in field.

\begin{figure}
\includegraphics[width=6.0cm]{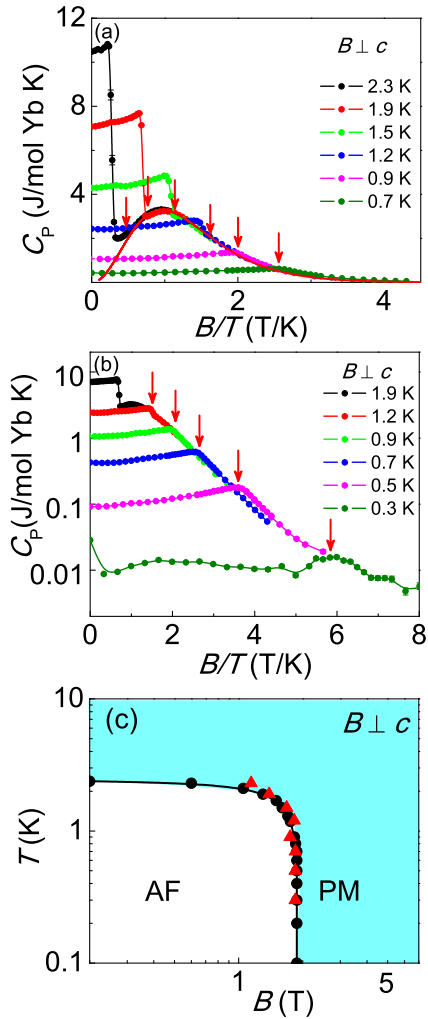}
\caption{(Color online) (a) The field dependencies of the specific heat $C_{\rm P}$ were obtained at different fixed temperatures, and were then plotted as functions of $B/T$. The red arrows indicate the onset of antiferromagnetic order, and the red line is the Schottky expression for the specific heat of a two level system with $g=2.5$. (b) An expanded view of the field dependencies of the  specific heat $C_{\rm P}$ measured from 0.3 K to 1.9 K. The red arrows indicate the onset of antiferromagnetic order. (c) The lowest temperature where $B/T$ scaling was observed in the specific heat $C_{\rm P}$ (red triangles) is virtually indistinguishable from the antiferromagnetic phase line $T_{\rm N}(B)$ (black circles) previously determined from specific heat measurements~\cite{wu2011}. The $B/T$ scaling is seen in the shaded region that extends over a very wide range of fields and temperatures where antiferromagnetic order is absent. \label{ScalingC}}
\end{figure}

The paramagnetic nature of the magnetic fluctuations also leads to $B/T$ scaling in the measured specific heat $C_{\rm P}$. The field dependence of $C_{\rm P}$ is plotted in Fig.~\ref{ScalingC}a for different fixed temperatures between 0.7 K and 2.3 K, and at lower temperatures in Fig.~\ref{ScalingC}b, where the field dependencies of $C_{\rm P}(B)$ are presented for 1.9 K$\geq T \geq 0.3$ K. For each temperature, $C_{\rm P}$ falls on an apparently universal function of $B/T$ above a characteristic value of $B/T$ marked by red arrows. Fig.~\ref{ScalingC}a shows the fields separating the scaling and nonscaling parts of the $C_{\rm P}(B)$ curves, and  the resulting curve closely resembles the phase line $T_{\rm N}(B)$ in Fig.~\ref{BT}. Like the magnetization $M$, the $B/T$ scaling evident in $C_{\rm P}$ betrays an underlying energy spectrum that has only two states. Accordingly, Fig.~\ref{ScalingC}a shows that $C_{\rm M}$ is well described in the paramagnetic phase by a Schottky expression, where the Zeeman splitting of the states $\Delta=g\mu_{\rm B}B$ with $g=2.5$.

\begin{figure}
\includegraphics[width=6.0cm]{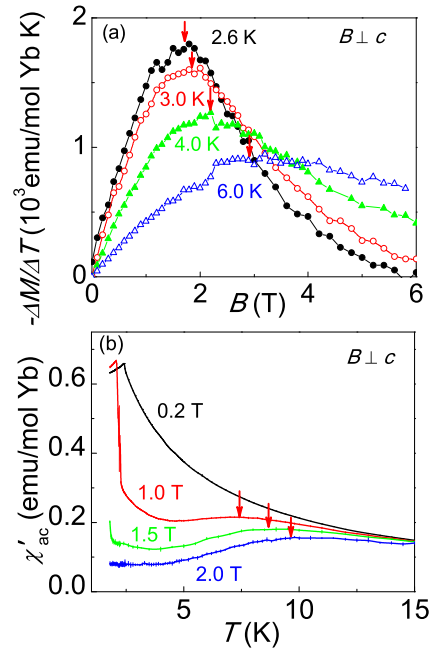}
\caption{(Color online) (a)  $-\Delta M/\Delta T \ vs \ B$ calculated as described in the text, for different fixed temperatures. (b) The temperature dependencies of the real part of the ac magnetic susceptibility $\chi^{\prime}_{ac}$, measured  at different fields. The red arrows in (a) and (b) mark the positions of maxima. \label{Mac}}
\end{figure}

The $B/T$ scaling that we have demonstrated in the field and temperature dependent resistivity $\rho$, magnetization $M$, and specific heat $C$ suggests that the predominant magnetic fluctuations that are present for $T\geq T_{\rm N}(B)$, and in the $T=0$ paramagnetic phase where $B$ exceeds the critical value of 1.85 T, are incoherent fluctuations of the Yb moments within their Zeeman split doublet ground state. Within the accuracy of our measurement, this single ion behavior extends to $T_{\rm N}$ itself, implying that critical fluctuations play a negligible role in $\YP$.  If this conclusion is correct, then the magnitude of the gap $\Delta$ between the Zeeman split ground state doublet of the Yb ions should provide the only energy scale for the paramagnetic part of the $\YP$ phase diagram. The importance of this energy scale near field-driven QCPs has recently been emphasized~\cite{hackl2011}.

\begin{figure}
\includegraphics[width=7.0cm]{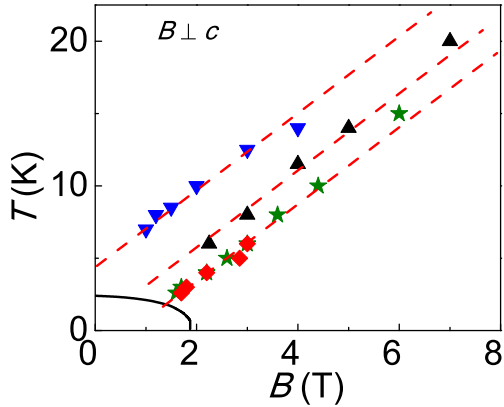}
\caption{(Color online) The field dependencies of the crossover temperatures determined from $-\Delta M/\Delta T$ ($T_{\rm M}$, red diamonds)), ac magnetic susceptibility ($T_{\chi}$, blue triangles), $d\rho/dB (T_{\rho}$, green stars), and the Zeeman energy $T_{\Delta}=\Delta/k_{\rm B}$ (black triangles) determined from the high temperature specific heat measurement. Solid line is the antiferromagnetic phase boundary $T_{\rm N}$(B) taken from Fig.~\ref{BT}. The dashed red lines are guides for the eye, indicating that the three different temperature scales have the same slope($\Delta T/\Delta B \simeq 2.6$ K/T). \label{BTcrossover}}
\end{figure}

The Zeeman gap $\Delta$ may be determined, in principle, from analyses of the magnetization $M$, resistivity $\rho$, and specific heat $C_{\rm P}$. The temperature derivative of the magnetization $\Delta M/\Delta T$ can be calculated from magnetization isotherms $M(B)$, measured at temperatures differing by $\Delta T = 0.05$ K according to $-dM/dT\simeq-\Delta M/\Delta T= -[(M(T+\Delta T, B)-M(T-\Delta T,B)]/(2 \Delta T)$. This procedure is repeated for a wide range of fields $B$, and the result is plotted in Fig.~\ref{Mac}a. We restrict ourselves here to temperatures $T\geq T_{\rm N}$. The field dependence of -$\Delta M/\Delta T$ displays a distinct maximum at a field $B_{\rm M}$ that moves to higher fields with increasing temperature. The temperatures $T_{\rm M}$ and fields $B_{\rm M}$ of the maxima in $-\Delta B/\Delta T$ are plotted in Fig.~\ref{BTcrossover}, where they are shown to be linearly related.

ac magnetic susceptibility measurements provide complementary information, since $\chi^{\prime}_{ac}$ is defined as the field derivative of the magnetization, measured as a function of temperature in different fixed dc fields (Fig.~\ref{Mac}b). When the dc magnetic fields are small, a sharp ordering anomaly is observed at $T_{\rm N}$, which passes out of our experimental temperature window $T\geq1.8$ K for $B\geq1$ T. In the  paramagnetic state at higher fields, $\chi^{\prime}_{ac}$ also has a maximum at $T_{\chi^{\prime}}$, which moves to larger temperatures with increasing fields. Fig.~\ref{Mac} shows that, like $T_{\chi^{\prime}}(B)$, $T_{\rm M}(B)$  increases linearly with magnetic field, at least for the limited range of fields where the magnetization and ac susceptibility measurements overlap. Intriguingly, the peak in $\chi^{\prime}_{ac}$ is not driven to $T=0$ as $B\rightarrow0$, but instead occurs at $\simeq 4.6$ K when $B=0$.

\begin{figure}
\includegraphics[width=6.0cm]{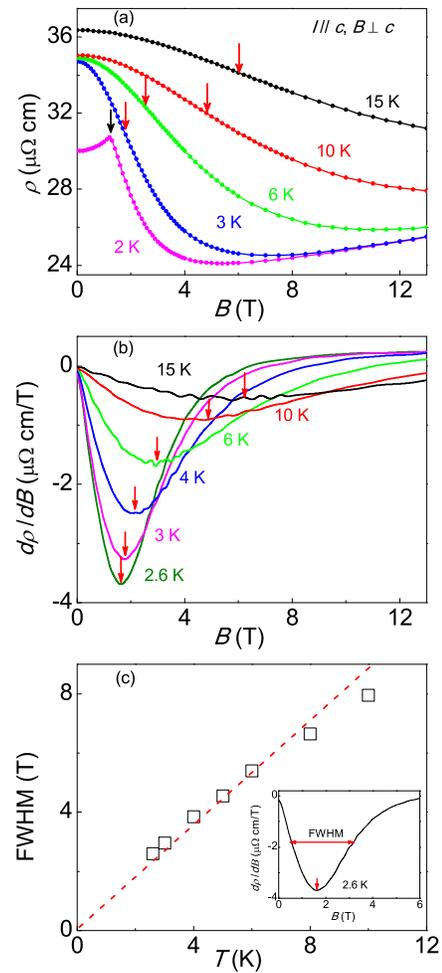}
\caption{(Color online) (a) Magnetic field dependencies of the electrical resistivity $\rho(B)$ measured at different fixed temperatures, as indicated. The black arrow indicates the antiferromagnetic transition, and red arrows  indicate the position of the negative peak in $d\rho/dB$ (b). (c) The full width at half maximum(FWHM) of the $d\rho/dB$ peak decreases linearly with decreasing temperature, and within the accuracy of our analysis extrapolates to zero for $T\rightarrow0$ (red dashed line). Inset: The FWHM is defined as the crossover width at half maximum (as indicated by the red horizontal arrow) of the $d\rho/dB$ peak, demonstrated here for $T=2.6$ K. \label{dRdB}}
\end{figure}

Since the resistivity and the magnetization are related for paramagnetic $\YP$, it follows that the field derivative of the resistivity $d\rho/dB$ will also have a peak that mirrors that of $\chi^{\prime}_{ac}=dM/dH$. The magnetoresistance of $\YP$ was measured for temperatures $T\geq T_{\rm N}$, as shown in Fig.~\ref{dRdB}a. The corresponding field derivative $d\rho/dB$ was determined numerically, and it is plotted in Fig.~\ref{dRdB}b. A negative maximum is found for $d\rho/dB$ that moves to larger fields with increased temperature. The fields $B_{\rho}$ and temperatures $T_{\rho}$ where $-d\rho/dT$ has its maximum should correspond to the fields $B_{\rm M}$ and temperatures $T_{\rm M}$  where $-\Delta M/\Delta T$ has its maximum.  Fig.~\ref{BTcrossover} confirms that $T_{\rho}$ and $T_{\rm M}$ are identical, within the accuracy of our analyses. The peak in $-d\rho/dT$ broadens markedly with increasingly temperature, and although the onset of antiferromagnetic order prohibits a direct measurement, its  full-width, half-maximum (FWHM)(Fig.~\ref{dRdB}c) extrapolates approximately to zero as $B\rightarrow0$.

\begin{figure}
\includegraphics[width=6.00cm]{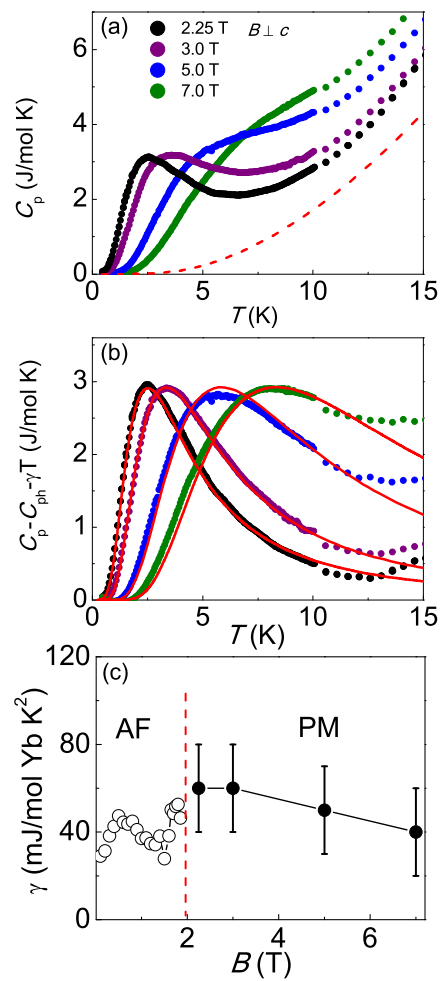}
\caption{(Color online) Temperature dependencies of the specific heat $C_{\rm P}$, measured at different fields. The red dashed line is the measured $B=0$ specific heat of the isostructural and nonmagnetic analog compound Lu$_{3}$Pt$_{4}$, which gives an estimate of the phonon contribution to the specific heat (see text). (b) The temperature dependencies of the specific heat after subtraction of the phonon contribution $C_{\rm Ph}$ and the electronic contribution $\gamma(B)T$. The solid lines are fits to the Schottky expression, described in the text. (c) The Sommerfeld coefficient $\gamma$ that was obtained from the fits in (b) is almost field independent. Vertical dashed line indicates $B_{\rm CEP} = 1.9$ T, where antiferromagnetic order vanishes. \label{CT}}
\end{figure}

The effect of Zeeman splitting on the ground doublet is most obvious in measurements of the temperature dependent specific heat $C_{\rm P}$, carried out in different fixed fields (Fig.~\ref{CT}a).  We separate $C_{\rm P}$ into two parts: $C_{\rm P}= C_{\rm M}+C_{\rm Ph}$. $C_{\rm Ph}$ is the contribution from the phonons, and we approximate this term by the specific heat measured in nonmagnetic but isostructural Lu$_{3}$Pt$_{4}$ (Fig.~\ref{CT}a). $C_{Ph}$ is taken to be field independent. $C_{\rm M}$ is the magnetic and electronic contribution to the specific heat, and we take $C_{\rm M}=\gamma(B)T + C_{\rm Schottky}$. $C_{\rm P}-C_{\rm Ph}-\gamma T$ is plotted in Fig.~\ref{CT}b, and indeed it consists of a peak that broadens and moves to higher temperatures with increasing field, much as we would expect for a Schottky contribution to the specific heat. Accordingly, we have fit  $C_{\rm P}- C_{\rm Ph} = \gamma(B)T + C_{\rm Schottky}$, where $C_{\rm Schottky}$ is the Schottky expression for two levels with equal degeneracy, separated by a gap $\Delta(B)$. The quality of these fits for fields from 2.25 T to 7 T is demonstrated in Fig.~\ref{CT}b. The Sommerfeld coefficient $\gamma(B)$ is approximately 40 mJ/mol-K$^{2}$ for $B=0$, and the minimal field dependence that is displayed in Fig.~\ref{CT}c likely reflects the inherent accuracy of our fits. $\gamma$ is always small, consistent with the apparent absence of any Kondo physics in $\YP$, and there is no evidence for any divergence of $\gamma$ at $B_{\rm CEP}$, in agreement with similar results on the resistivity coefficient $A$ (Fig.~\ref{dRdT}b).

Fig.~\ref{CT}b shows that $C_{\rm M}=C_{\rm P}-\gamma T-C_{\rm Ph}$ is well fitted by the Schottky expression for fields from 2.25 T to 7 T, and the field dependence of the temperature scale $T_{\Delta}=\Delta/k_{\rm B}$ that results from these fits has been added to Fig.~\ref{BTcrossover}. As expected, $T_{\Delta}$ increases linearly with field.  While the temperature scales $T_{\rm M}$, $T_{\rm \chi^{\prime}}$, $T_{\rm \rho}$, and $T_{\rm \Delta}$ are not all identical, in each case we find that their slopes $\Delta T/\Delta B\simeq 2.6$ K/T (Fig.~\ref{BTcrossover}), which is also  consistent with the value $g=2.5$ found in the scaling of the specific heat at very low temperatures (Fig.~\ref{ScalingC}a). It is tempting to believe that all these scales originate with the Zeeman splitting of the Yb doublet ground state.

\section{Discussion and Conclusion }

Our measurements suggest that $\YP$ is a particularly simple system. Throughout the paramagnetic phase $T\geq T_{\rm N}$ and $B\geq B_{\rm CEP}$, the magnetic and electronic specific heat $C_{\rm M}(T,B)$, the magnetization $M(T,B)$, and even the resistivity $\rho(T,B)$ are all dominated by strong magnetic fluctuations, where the only characteristic energy scale results from the Zeeman splitting of an energetically isolated, Yb doublet ground state.  These single ion, paramagnetic fluctuations extend down to $T_{\rm N}(B)$ itself, indicating that critical fluctuations are always very weak. This may reflect the fact that the N\'{e}el state vanishes at $B_{\rm CEP} = 1.85$ T in a field-driven critical endpoint, much as is found for antiferromagnetic insulators~\cite{stryjewski1977,birgenau1972,dillon1978}. Quantum critical fluctuations are still possible, in principle, if this transition is  weakly first order. We speculate that the absence of these quantum critical fluctuations in Yb$_{3}$Pt$_{4}$ may result from  an inherent mean-field like character that is evident in the phase line T$_{N}$(B), from the B=0 order parameter found in neutron diffraction measurements~\cite{janssen2010},and in the appearance of the specific heat transition itself~\cite{wu2011}.  The highly localized character of the moments in Yb$_{3}$Pt$_{4}$ prohibits the sorts of quantum critical fluctuations between states with different Fermi surface volumes that were reported in YbRh$_{2}$Si$_{2}$, suggesting that they may be a larger part of the quantum critical fluctuations of the more hybridized heavy fermions than was previously appreciated.

$\YP$ is a metal, and the near-constancy of the Sommerfeld coefficient for fields both larger and smaller than $B_{\rm CEP}$ suggests that there is a Fermi liquid state that underlies both the antiferromagnetic and paramagnetic phases in $\YP$. The $T^{2}$ temperature dependence of the electrical resistivity is only observed when antiferromagnetic order disables the paramagnetic fluctuations, suppressing the spin-disorder scattering that otherwise obscures the Fermi liquid component of the resistivity. The smallness of the Sommerfeld coefficient indicates that the exchange coupling of the conduction electrons to the Yb moments is weak, and that the quasiparticle mass enhancement is minimal.
It is fair to say that the Fermi liquid in $\YP$ simply coexists with the Yb moments, and that it is almost unaffected by the onset of antiferromagnetic order. $\YP$ seems to have much more in common with  elemental rare earth metals like Gd or Dy, where magnetic order occurs well above the extremely low or even vanishing temperature scales where Kondo physics could play a role, than heavy fermions like $\YRS$, where the Kondo effect is largely complete by the time magnetic order is established.

Our measurements provide definitive answers to the questions that we posed in the introduction.

\noindent$\bullet$ Is non-Fermi liquid behavior found near the $T_{\rm N}=0$, $B=B_{\rm CEP}=1.85$ T critical endpoint? Given the first order character of this transition, quantum critical fluctuations are weak, at best. We have showed that paramagnetic fluctuations of individual Yb moments dominate all measured quantities down to the antiferromagnetic phase line itself. non-Fermi liquid behaviors such as \mbox{$\Delta\rho= BT^{1+\delta}$} are entirely absent near $B_{\rm CEP}$.

\noindent$\bullet$Is a heavy Fermi liquid found once magnetic fields suppress antiferromagnetic order? A Fermi liquid underlies both the antiferromagnetic and paramagnetic phases of $\YP$, but the Sommerfeld coefficient is small in both, signalling a small quasiparticle mass enhancement.  There is no sign of Fermi liquid breakdown in paramagnetic $\YP$, signalled in other systems by divergencies of the Sommerfeld coefficient $\gamma$ or the resistivity coefficient A as the field approaches $B_{\rm CEP}$ from above.

\noindent$\bullet$Is there any indication of electronic delocalization in $\YP$? $\YP$ appears to be an extreme case of moment localization. Outside the range of fields and temperatures where antiferromagnetic order is stable, the  electrical resistivity, magnetization, and specific heat all display the $B/T$ scaling that is expected for decoupled and fully incoherent magnetic moments, where the spacing between the underlying energy levels increases linearly with magnetic field. The ubiquity of $B/T$ scaling suggests that these levels originate with the  well-separated doublet ground state in $\YP$, which is Zeeman split in field. This single ion behavior dominates in the absence of antiferromagnetic order, suggesting that the Yb moments are always localized, seemingly ruling out the possibility of electronic delocalization and an expansion of the Fermi surface at $T_{\rm N}$, as is found in systems like $\YRS$.

It is interesting to consider how the rather minimal physics of localized $\YP$ might be connected to the rich physics that is found in heavy fermions with bona fide QCPs.  Is there a generalized $T=0$ phase diagram that can accommodate both?  We present a phase diagram in Fig.~12 that proposes just such a  connection. Since this proposed phase diagram is based largely on experimental results in Yb$_{3}$Pt$_{4}$, and further experimental investigation will be required to establish whether it may have more universal application. One axis of this phase diagram is inspired by the Doniach phase diagram~\cite{doniach1977}, and represents the degree of hybridization $\Gamma$  between the moment-bearing f-electrons and conduction electrons. Applied pressure increases $\Gamma$ for Ce compounds, but decreases $\Gamma$ for Yb compounds~\cite{thompson1994,goltsev2005}.  The Doniach argument associates magnetic order arising from the Rudermann-Kittel-Kasuya-Yosida (RKKY) interaction with weak hybridization, although the increasing influence of Kondo physics ultimately leads to its suppression at a QCP for a critical value of $\Gamma=\Gamma_{\rm QCP}$.  The second axis of this $T=0$ phase diagram is magnetic field, which generally suppresses antiferromagnetic order. Antiferromagnetic order is stable at  $T=0$ when $B\leq B_{\rm N}(\Gamma)$ and for $B=0$,  when $\Gamma\leq\Gamma_{\rm QCP}$.

\begin{figure}
\includegraphics[width=7cm]{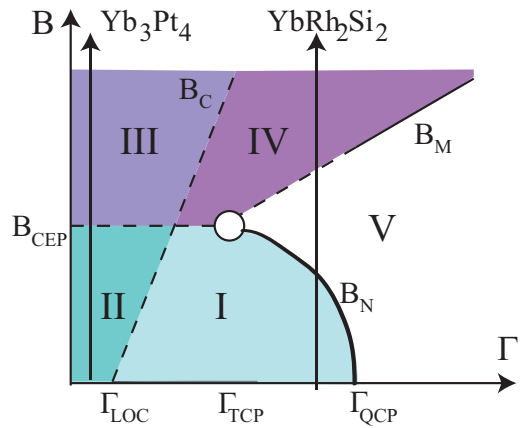}
\caption{(Color online) Field $B$ - hybridization $\Gamma$ phase diagram for $T=0$. The antiferromagnetic phase line $B_{\rm N}$($\Gamma$) has a continuous region that terminates for $B=0$, $\Gamma=\Gamma_{\rm QCP}$ (solid line) and a first order part (dashed line) that terminates at $\Gamma\rightarrow0$, $B_{\rm CEP}$,  separated by a tricritical point (White circle, $\Gamma=\Gamma_{\rm TCP}$). Regions I and II are antiferromagnetically ordered, regions III,IV, and V are not.  Dashed line $B_{\rm C}$($\Gamma$) separates regions II and III, having localized Yb moments, from Regions I, IV, and V, where there are differing degrees of electronic localization (see text). The line $B_{\rm M}$($\Gamma$) separates regions IV (light mass Fermi liquid) from region V (heavy mass Fermi liquid). It is not known where $B_{\rm M}$($\Gamma$) intersects the antiferromagnetic phase line $B_{\rm N}$($\Gamma$) (dashed line). The evolution of the $T=0$ states of $\YP$ and $\YRS$ with increasing field is  indicated by vertical arrows.  \label{BTSketch}}
\end{figure}

The persistence of field-temperature scaling for compounds that are tuned to the vicinity of the ($B=0$, $\Gamma=\Gamma_{\rm QCP}$) QCP suggests that the phase line $B_{\rm N}(\Gamma)$ is second order for an appreciable range of the hybridization parameter $\Gamma$, terminating for $B=0$ at $\Gamma=\Gamma_{\rm QCP}$~\cite{custers2003}.  $\YRS$ forms very close to $\Gamma_{\rm QCP}$, and the fragility of its antiferromagnetic state is evident from both the tiny ordered moment~\cite{ishida2003} and by the small amounts of doping that are required to drive $T_{\rm N}\rightarrow0$~\cite{custers2003,custers2010}.  Larger chemical pressures are responsible for the absence of antiferromagnetic order in YbIr$_{2}$Si$_{2}$, which can be restored by the subsequent application of hydrostatic pressure~\cite{yuan2006}. High pressures are expected to stabilize antiferromagnetic order at progressively higher fields, an effect that is reproduced by Co-doping in Yb(Rh$_{1-x}$,Co$_{x}$)$_{2}$Si$_{2}$~\cite{friedemann2009}. A different behavior is found in compounds like $\YP$, where the exchange coupling $\Gamma$ is very small ($\Gamma\rightarrow0$) and the field-driven phase transition $T_{\rm N}=0$ is first order. The magnetic fields required to suppress antiferromagnetic order to $T_{\rm N}=0$ form a line of $T=0$ transitions that emanate from a tricritical point with $\Gamma_{\rm TCP}$ that separates this first order part of the $B_{\rm N}(\Gamma)$ phase line with $\Gamma\rightarrow0$ from the continuous regime with $\Gamma\rightarrow\Gamma_{\rm QCP}$~\cite{misawa2008,misawa2009}. There is some initial evidence that the antiferromagnetic ground state is achieved via a first-order transition in Co-doped $\YRS$~\cite{klingner2010}, suggesting that it may be  possible to span this tricritical point with an appropriate combination of magnetic fields and chemical pressure.

Very different types of electronic behaviors are found in the different regimes of this $T=0$ phase diagram. All these $f$-electron based compounds start with the same high temperature state, where spatially localized moments fluctuate independently and are essentially decoupled from the conduction electrons. With lowered temperature, magnetic order and Kondo compensation compete to determine the final $T=0$ state. In systems like $\YP$, $T_{\rm N}$ is larger than $T_{\rm K}$, and so the ground state is magnetic order of spatially localized moments, where the related $f$-electrons or holes are excluded from the Fermi surface. Magnetic fields suppress the $T=0$ antiferromagnetic order in $\YP$, and the robust $B/T$ scaling in the paramagnetic regime indicates that the localized moments persist, creating a paramagnetic state that is stable even for $T=0$. In $\YRS$, $T_{\rm K}$ is much larger than $T_{\rm N}$. Here, the Yb-based $f$-holes and the conduction electrons are strongly entangled, with both contributing to the Fermi surface of the $T=0, B=0$ ordered state. Here, too, magnetic fields suppress antiferromagnetic order~\cite{gegenwart2002}, but the transition in $\YRS$ is accompanied by an expansion of the Fermi surface that produces a heavy Fermi liquid~\cite{paschen2004,hackl2011}. A second transition or crossover is found at $B_{\rm M}\simeq10$ T~\cite{tokiwa2005}, which is accompanied by a broadened step in the magnetization and a step like reduction  in the Sommerfeld constant, suggesting the formation of a new Fermi liquid with substantially reduced quasiparticle mass and interactions~\cite{gegenwart2006}. High pressure measurements on $\YRS$ find that $B_{M}$ decreases with increasing pressure (decreasing $\Gamma$) as indicated in Fig.~12. This general trend has been reported as well in a number of different heavy fermion and mixed valence compounds~\cite{hirose2011}. The exact nature of the transition or crossover at $B_{\rm M}$ remains uncertain. de Haas - van Alphen measurements~\cite{rourke2008} support the proposal
 that a Lifshitz transition occurs in $\YRS$ at $\simeq$10 T, where the majority spin sheet of the Fermi surface vanishes to produce a more weakly correlated Fermi liquid~\cite{kusminsky2008}. Electronic structure calculations suggest instead a gradual crossover that is driven by Zeeman splitting of the quasiparticle states, a process that redistributes spectral weight among bands with different masses, while leaving the number of states contained by the Fermi surface unchanged between the light and heavy Fermi liquid states~\cite{zwicknagl2011}.  Neither scenario suggests that there is an actual localization of the f-holes at $B_{\rm M}\simeq 10$ T.

The complete destruction of the heavy fermion state is projected to occur at a much higher field $B_{\rm C}$~\cite{kusminsky2008}, resulting in a high field state where the Yb moments and the conduction electrons are decoupled. The definitive absence of heavy fermion character in $\YP$, where the Yb moments and the conduction electrons are nearly decoupled, prompts our suggestion (Fig.~12) that a smaller field is required to suppress the heavy fermion state as $\Gamma$ decreases, ultimately producing a $B=0$ state with $\Gamma\leq\Gamma_{\rm LOC}$ where moments are always localized.  We note that such a transition has been observed in  $\YRS$, where a pressure $P\simeq10$ GPa causes the $B=0$ ordering transition become first order~\cite{plessel2003}, and the ordered Yb moment increases dramatically from 0.02 $\mu_{B}$/Yb at 1 bar~\cite{ishida2003} to $\sim1.9 \mu_{B}$/Yb at 16.5 GPa~\cite{plessel2003}.  The latter value is similar to the $B=0$ moment found in $\YP$, which is in turn close to the expected value for a Yb doublet ground state when $T_{\rm K}\rightarrow0$, signalling that the Yb moments have become largely decoupled from the conduction electrons. These data suggest that $B_{\rm C}$($\Gamma$) intersects the $B=0$ axis at $\Gamma_{\rm LOC}\leq\Gamma_{\rm TCP}$. Understanding how the $B_{\rm C}$($\Gamma$) line passes through the antiferromagnetic phase and connects to a  $B=0$ moment localization transition will require challenging new measurements that use high pressures or chemical pressure to drive localization, with the subsequent addition of magnetic fields to drive the resulting $T=0$ transition towards the $B_{\rm N}$($\Gamma$) phase line itself.

\begin{acknowledgments}
The authors acknowledge valuable assistance from M. S. Kim and K. Park. Work at Stony Brook University is supported by the National Science Foundation under grant DMR-0907457.
\end{acknowledgments}


\end{document}